\documentclass[a4paper,conference]{IEEEtran}
\IEEEoverridecommandlockouts

\usepackage{cite}
\usepackage{amsmath,amssymb,amsfonts}
\usepackage{algorithmic}
\usepackage{graphicx}
\usepackage{textcomp}
\usepackage{xcolor}
\def\BibTeX{{\rm B\kern-.05em{\sc i\kern-.025em b}\kern-.08em
    T\kern-.1667em\lower.7ex\hbox{E}\kern-.125emX}}
\begin{document}

\title{GDPRShield: AI-Powered GDPR Support for Software Developers in Small and Medium-Sized Enterprises \thanks{This work has been submitted to the EuroUSEC 2025 for possible publication. Copyright may be transferred without notice, after which this version may no longer be accessible.}}

\author{\IEEEauthorblockN{Tharaka Wijesundara}
\IEEEauthorblockA{\textit{School of Computing Technologies
} \\
\textit{RMIT}\\
Melbourne, Australia \\
s4063322@student.rmit.edu.au}
\and
\IEEEauthorblockN{Mathew Warren}
\IEEEauthorblockA{\textit{Research \& Innovation} \\
\textit{RMIT}\\
Melbourne, Australia \\
matthew.warren2@rmit.edu.au}
\and
\IEEEauthorblockN{Nalin Arachchilage}
\IEEEauthorblockA{\textit{School of Computing Technologies} \\
\textit{RMIT}\\
Melbourne, Australia \\
nalin.arachchilage@rmit.edu.au}
}

\maketitle

\begin{abstract}
With the rapid increase in privacy violations in modern software development, regulatory frameworks such as the General Data Protection Regulation (GDPR) have been established to enforce strict data protection practices. However, insufficient privacy awareness among SME software developers contributes to the organization's failure to comply with GDPR. For instance, a developer unfamiliar with data minimization principles may accidentally build a system that gathers too much information, which can break GDPR rules and lead to fines. One reason for this lack of privacy awareness is that developers in SMEs often take on multidisciplinary roles (e.g., front-end and back-end development, database management, and privacy compliance), which does not ensure specialization or expertise in privacy. This lack of privacy awareness of software developers may lead to poor privacy attitudes, which ultimately may hinder the development of a strong organizational privacy culture. However, if SMEs are able to achieve GDPR compliance, they may gain competitive advantages—such as increased user trust and marketing value—for their software applications and their organization, especially when compared to other SMEs that do not comply with GDPR.

Therefore, in this paper, we introduce a novel AI-powered framework called "GDPRShield," which is specifically designed to enhance the GDPR awareness of software developers in SMEs and, through this awareness, enhance their privacy attitudes. At the same time, GDPRShield improves the motivation of software developers to comply with GDPR from the early stages of software development. GDPRShield leverages functional requirements written as user stories to enhance software developers' GDPR awareness by providing them comprehensive GDPR-based privacy descriptions tailored to each requirement. At the same time that GDPR awareness is improved, GDPRShield boosts developers' motivation to comply with GDPR by providing the real-world consequences of GDPR noncompliance, such as heavy fines, reputational damage, and loss of user trust, aligning them with functional requirements. This motivation will lead software developers to use GDPRShield to enhance their GDPR awareness and comply with GDPR. This improved awareness will lead to improving privacy attitudes among software developers, which will eventually help SMEs strengthen their privacy culture in the long run.

\end{abstract}


\section{Introduction}
\label{intro}
With the rising concerns of privacy violations \cite{sobers2024}, governments worldwide have introduced various data protection regulations \cite{unctad2024}, such as the General Data Protection Regulation (GDPR) in the European Union (EU) \cite{gdprEU}, to ensure strict data protection practices \cite{PrybyloMaxwell24}. These regulations mandate that organizations \footnote{from this point onwards, we refer to organizations in the software development domain} comply with these privacy requirements and highlight that non-compliance will lead to financial penalties \cite{PrybyloMaxwell24, JainVijayanta23}. For instance, in December 2024, Italy’s data protection authority fined OpenAI €15 million for violations related to its ChatGPT platform \cite{openaiGarante2024}. OpenAI used personal data to train ChatGPT without asking for permission \cite{openaiGarante2024}. It exemplifies that simply having regulations in place in the region where the organization operates does not guarantee a successful implementation of privacy requirements within the organization \cite{PrybyloMaxwell24}. Instead, a strong privacy culture within the organization and positive privacy attitudes in software development teams ensure that privacy requirements are properly implemented in software development \cite{PrybyloMaxwell24, AyalonOshrat17, IritHadarTomer18, TahaeiMohammad21}, and it was found in some surveys as well \cite{TahaeiMohammad21, PrybyloMaxwell24}. With this, one could argue that there is a strong relationship between the organization's privacy culture and the software development teams' privacy attitudes. 

This relationship may be critical for Small and Medium-sized Enterprises (SMEs). In SMEs, software developers \footnote{from this point onwards, software developers will refer to those who work in SMEs unless explicitly specified} who are responsible for implementing privacy requirements \footnote{the reference highlights that an employee within an SME who is assigned to multiple roles may be responsible for cybersecurity, which includes privacy \cite{enisa2019cybersecurity}.} often lack proper privacy knowledge \cite{BarlettaVitaDesolda22, HjerppeKalle19, LiZeWerner22} including technical privacy proficiency \cite{mbego2025data}. For example, in general, many SMEs struggle even to encrypt user data \cite{gdpr_eu_2019SME}. This is supported by a survey conducted in 2019 with small organizations in Europe (which can be considered a subset of SMEs), as it indicates that 22\% of them do not use any technical measures (e.g., encryption or pseudonymization) to protect personal data \cite{gdpr_eu_2019SME}. This may be partly due to the fact that the software developers in SMEs are typically multi-disciplined \footnote{software developers in SMEs engage not only in software development but also in activities such as requirement specification, software design, and testing}, taking on various roles and responsibilities, not only in privacy \cite{RivasLornel08, enisa2019cybersecurity}. As a result, software developers in SMEs have less privacy expertise, which could potentially lead to weaker privacy attitudes. It was demonstrated by the results of some surveys, which revealed that in SMEs, software developers with varying privacy attitudes exist \cite{TahaeiMohammad21}. They may interpret others' privacy expectations through the lens of their own beliefs, assuming that others should have similar views on privacy \cite{TahaeiMohammad21}, which may affect the organization's overall privacy culture. For example, some developers may have strong positive privacy attitudes, such as "I always put myself in the other person’s shoes. I would not like my data to be tampered with." \cite{TahaeiMohammad21}. On the other hand, someone may have negative privacy attitudes, such as "I have nothing to hide," implying that others' privacy does not need to be considered \cite{TahaeiMohammad21}. These negative privacy attitudes are barriers to implementing privacy in software development, which results in a bad organizational privacy culture \cite{TahaeiMohammad21}. Hence, improving privacy attitudes is essential, especially in SMEs, where weak privacy expertise directly contributes to poor attitudes and behaviors around the privacy of software developers. 


SMEs specifically face difficulties in implementing necessary GDPR requirements because of the lack of privacy expertise in SMEs \cite{WaidelichLukas23, gdpru2019survey}. For example, understanding the legal language of GDPR may be hard for someone who does not have privacy expertise. Also, they do not outsource this task due to its high cost \cite{WaidelichLukas23, gdpru2019survey}, which is unaffordable for SMEs \cite{LiZeWerner22, Sirur2018AreWT}. A survey conducted in Ireland in 2023 found that 66\% of organizations (many of them small and medium-sized organizations) thought that following GDPR rules turned out to be more expensive than they had expected back in 2018 (i.e., when GDPR first came in) \cite{ForvisMazars2023GDPRSurvey}. Further, the European Commission acknowledged in its official evaluation that GDPR is challenging, especially for small and medium-sized enterprises \cite{EU2020GDPRReview}. This non-compliance makes SMEs more vulnerable to non-compliance penalties (i.e., monetary penalties). However, it also presents a clear opportunity (benefit) for SMEs: if SMEs can comply with GDPR, they gain competitive advantages, such as improved user trust and promotional value (i.e., marketing advantage) for their software applications and their organization \cite{emarketer2025} among other SMEs who do not comply with GDPR. Since they have additional benefits of complying with GDPR, if their GDPR awareness can be improved, they may tend to apply it in practice. Thus, the aim of our research is threefold: first, to increase software developers' GDPR awareness; second, to improve their privacy attitudes through this awareness; and third, to increase their motivation to act on that awareness (i.e., comply with GDPR). Without motivation, even if awareness improves, developers may not apply it in practice. Further, this awareness may help to sustain a strong privacy culture within the organization in the long run, since an organization with a strong privacy culture attracts software developers with strong positive privacy attitudes \cite{TahaeiMohammad21}.


Considering everything, we proposed a novel AI-powered framework called "GDPRShield." GDPRShield aims to improve the GDPR awareness of software developers in SMEs and, through this awareness, enhance their privacy attitudes. To do that, the framework provides software developers with detailed GDPR-related privacy descriptions for each functional requirement (i.e., those written as user stories), including how to comply with GDPR, why that compliance is needed, and from which part of the GDPR it was extracted. In addition to awareness, GDPRShield also aims to improve the motivation of software developers to comply with GDPR. To do that, the framework provides real-world consequences of GDPR noncompliance by relating them to functional requirements. These noncompliance cases illustrate what could happen if the GDPR requirements are not met.


This paper is structured as follows: we discuss the related work in Section \ref{related_work} and highlight the research gap to where this paper is proposed; in Section \ref{prop_fw}, we present the proposed framework in detail with its features and how they work; in Section \ref{usecase}, we present a real-world use case of the framework; and finally, in Section \ref{concl_fw}, we present the contribution of the framework as a summary and potential future work. 

\section{Related Work}
\label{related_work}

Due to a lack of privacy knowledge, expertise, and financial resources within the organization, SMEs are unable to comply with GDPR regulations \cite{LiZeWerner22, Sirur2018AreWT, FreitasMaria18, PedrosoLuís21, WaidelichLukas23, AlhirabiNada23, RivasLornel08}. To help SMEs comply with GDPR, Li et al. \cite{LiZeWerner22} proposed a list of privacy requirements derived from GDPR to simplify its legal jargon as well as a tool to automatically check whether the system complies with the regulation. Similarly, Brodin \cite{BrodinMartin19} proposed a structured, three-phase (i.e., analysis, design, and implementation) framework by breaking down GDPR compliance into actionable steps, ensuring SMEs can systematically address GDPR requirements. Even though both solutions are specifically aimed at helping SMEs (i.e., those with less privacy expertise) comply with GDPR, some level of privacy knowledge is required to use these solutions. For instance, in Li et al.'s solution \cite{LiZeWerner22}, selecting the required GDPR requirements for the application from the list of requirements requires basic privacy awareness. Similarly, during the design phase of Brodin's \cite{BrodinMartin19} solution, a basic understanding of GDPR is required to create or update GDPR-compliant policies. Additionally, Li et al. \cite{LiZeWerner22} dropped some principles of GDPR, such as lawfulness and fairness, making it a less comprehensive solution for GDPR awareness. Therefore, these solutions may not address SMEs' one of the major problems, which is a lack of GDPR awareness.

Further, to specifically generate privacy requirements in the agile context based on user stories and to detect privacy-related information in user stories, Herwanto et al. \cite{HerwantoGuntur24} and Casillo et al. \cite{CasilloFrancesco22} proposed Natural Language Processing (NLP)-based solutions, respectively. Even though Herwanto et al.'s \cite{HerwantoGuntur24} solution partially covers GDPR principles when generating privacy requirements, it does not provide comprehensive GDPR awareness for developers because it relies on predefined privacy requirement templates with placeholders filled by extracting relevant information from user stories (e.g., subject of the story, type of user data, etc.). These templates may not be comprehensive enough for developers with limited privacy knowledge to understand GDPR comprehensively. Similarly, although Casillo et al.'s \cite{CasilloFrancesco22} solution helps developers identify whether a user story requires privacy consideration (i.e., by checking for privacy-related information), it is not based on GDPR requirements, implying that it does not provide GDPR awareness to developers. Because of the above issues, even though these solutions help developers consider privacy during the requirement specification stage, they do not provide comprehensive GDPR awareness, implying that SMEs with limited privacy knowledge may not benefit from these solutions.

To summarize, these approaches may not provide solutions to the problem of developers in SMEs avoiding GDPR compliance. As we discussed in the introduction, one of the reasons for avoiding GDPR compliance is a lack of privacy knowledge \cite{BarlettaVitaDesolda22, HjerppeKalle19, LiZeWerner22}, which may negatively affect software developers' privacy attitudes \cite{PrybyloMaxwell24, AyalonOshrat17, IritHadarTomer18, TahaeiMohammad21}. At the same time, these solutions do not aim to motivate developers to comply with GDPR. Therefore, the above-discussed existing solutions will not help improve software developers' GDPR awareness, which may be critical to enhancing their privacy attitudes and the overall privacy culture within the organization. As a solution, we propose GDPRShield, which aims to improve software developers' GDPR awareness and, through that, enhance their privacy attitudes while also motivating them to comply by presenting real-world noncompliance issues.

\section{Proposed Framework}
\label{prop_fw}

\begin{figure*}[h]
    \centering
    \includegraphics[width=\textwidth]{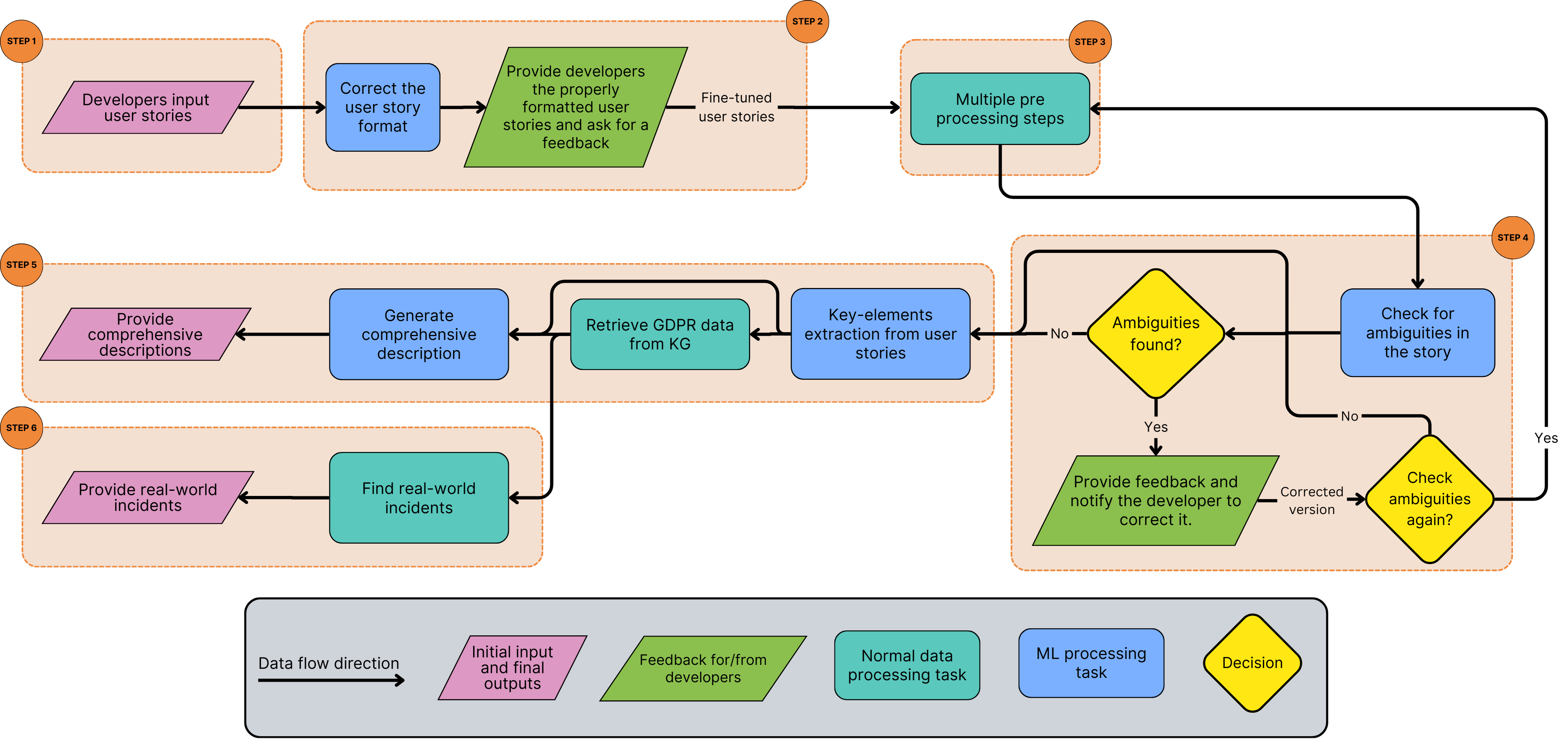}
    \caption{Proposed framework for GDPR-based comprehensive privacy description generation and real-world non-compliance incident suggestion}
    \label{fig:fw-diagram}
\end{figure*}

Figure \ref{fig:fw-diagram} presents the proposed framework, which is designed to help software developers improve their GDPR awareness through GDPR-based comprehensive privacy descriptions generated based on the information in user stories and to provide developers with real-world non-compliance incidents to motivate them towards implementing GDPR requirements. This awareness may eventually affect software developers' privacy attitudes, as there is a correlation between them, as we argued in the introduction, and eventually help to improve the organizational privacy culture. 

The framework flows through multiple steps. First, developers are required to specify functional requirements as user stories and input them into the framework. GDPRShield then checks the user stories for grammar, spelling, and formatting issues. Next, it identifies which functional requirements require GDPR compliance. Following that, it detects and highlights ambiguities in the user stories (discussed in detail in Section \ref{amb_identify} and Appendix \ref{sec-apndx-ambi}) and requests developers to resolve them. Once the ambiguities are addressed, the framework automatically generates comprehensive GDPR-related privacy descriptions for each user story. It then presents real-world GDPR non-compliance cases. Finally, to assess privacy attitudes at any time, developers are provided with a questionnaire to be completed using a Likert scale.

The following sections will discuss in detail how the framework improves GDPR awareness by generating GDPR-related privacy descriptions for each user story (Section \ref{priv_desc}), how GDPR-based non-compliance cases are shown to enhance developers’ motivation to comply with GDPR (Section \ref{real-world}), and finally, how the privacy attitude of software developers is evaluated (Section \ref{priv_attitude_eval}), which contributes to the overall privacy culture. 


The input for the framework, as shown in Figure \ref{fig:fw-diagram} (Step 1), is user stories that have a widely used specific format that consists of the who, what, and why of a requirement \cite{LucassenGarm16, CasilloFrancesco22, AMNA2025112357}. Here, "who" refers to the actor of the requirement, "what" refers to the action that the actor does, and "why" refers to the reason why the action is taken. For example, "As a user (who), I want to register an account on the system by providing my personal information (what), so that I can use the platform (why)." However, developers may make spelling and grammar errors, as well as fail to follow the well-known structure of user stories when writing them. Proceeding with these issues may cause inaccurate results in the next steps, as we rely on NLP tasks \cite{VadlapatiPraneeth23}. As a result, as the initial step of step 2 (Figure  \ref{fig:fw-diagram}), we propose using a Large Language Model (LLM) to detect and correct grammar and spelling errors, as well as the format of user stories, assuming that user stories written by developers contain the who, what, and why of user stories, even if the format differs. To achieve this, we can use OpenAI’s GPT-4 (via APIs) \cite{openaiGPT42025} with a specific prompt written for this task. For example, a prompt similar to this, \textit{"You will be given a user story that may contain grammar mistakes, spelling errors, and may not follow the standard user story format. Your task is to correct all grammar and spelling mistakes and rewrite the user story in the standard format "As a [who], I want to [what] so that [why]." If any of these elements (who, what, or why) are missing or cannot be identified, do not assume anything; instead, give a warning message stating which element is missing."} can be used. Furthermore, as AI solutions are not 100\% accurate, we propose receiving developer feedback through an interface at this stage to fine-tune the corrected user stories if the LLM produced any inaccurate results. 



Once the user stories have been properly formatted and all spelling and grammar errors have been corrected, they should be fed into the Machine Learning (ML)-powered operations (e.g., LLM operations) for the next steps, as shown in Figure \ref{fig:fw-diagram} (Steps 4-6). However, before feeding them to ML operations, we need to transform those identified user stories that are in raw text format into a structured, clean format suitable for further processing. To achieve that, we suggest pre-processing the user stories as the next step of the framework (Figure \ref{fig:fw-diagram} (Step 3)). 

\subsubsection{Pre-processing}
First, the user stories are passed through a tokenizer, which splits the text into a set of tokens. For example, we can use the WordPiece tokenizer \cite{JacobDevlinBERT, YonghuiWuWordPiece16}.

Next, we propose using contextual word embedding to convert tokens into numerical vectors \cite{HerwantoGuntur24, AzeemMuhammad24}, since raw text cannot be directly fed into ML models. For word embedding, we suggest using a transformer-based model, such as BERT (Bidirectional Encoder Representations from Transformers) \cite{JacobDevlinBERT} or RoBERTa \cite{LiuRoBERTa}, since BERT and RoBERTa have been shown to produce high-quality embeddings that improve the performance of NLP tasks \cite{HerwantoGuntur21, AzeemMuhammad24}. These models, along with their native tokenizers, are readily available in Hugging Face’s Transformers library \cite{HuggingFace_Transformers}.


\subsubsection{Ambiguities identification}
\label{amb_identify}
Next, we need to consider the ambiguities associated with user stories because even if user stories are written following a specific format and are free of spelling and grammar errors, they may contain other types of ambiguities (as described in Table \ref{table:ambiguities} in Appendix \ref{sec-apndx-ambi}), making them difficult to understand \cite{HerwantoGuntur24, AmnaAnis22, AMNA2025112357, DALPIAZ20193, AMNA2022106824}. If we proceed with these ambiguities, it may lead to non-precise results because of non-precise user stories. For example, "user locations" (i.e., the example under lexical ambiguity in Table \ref{table:ambiguities} in Appendix \ref{sec-apndx-ambi}) is a vague term, and it may refer to home, general area, city, etc. However, the original user story can be refined so that it clearly communicates to which user location it refers: \textit{"As a delivery driver, I want to access users’ real-time GPS coordinates during delivery hours so that I can complete deliveries efficiently."}. The latter explained the sensitivity of information more precisely than the former, which may affect the privacy descriptions. In the ambiguous user story, the GDPR articles that can be referenced are Article 6 (lawful basis for processing) and Article 5(1)(c) (data minimization) as general GDPR articles, since it is unclear what type of location data is used by the system. In the refined story, it is clearly mentioned that the real-time GPS coordinates are used by the system. In that case, it can be considered personal data under Article 4(1) and may require additional safeguards under Article 25 (data protection by default), given the increased identifiability and risk of continuous tracking. Therefore, to overcome these ambiguity issues, we propose detecting ambiguities in user stories as shown in Figure \ref{fig:fw-diagram} (Step 4) before proceeding to the next steps. 

To achieve this, we suggest getting insights from ambiguity detection frameworks such as the Quality User Story (QUS) framework \cite{LucassenGarm15QUS} and the Ambiguity Tracking and Resolution for User Stories (AmbiTRUS) framework \cite{AMNA2025112357} and incorporating NLP techniques, as they have shown positive results in previous ambiguity detection attempts such as the Automatic Quality User
Story Artisan (AQUSA) tool \cite{LucassenGarm16AQUSA, LucassenGarm15QUS}, and the NLP-based lexical ambiguity detection tool \cite{DALPIAZ20193}. Further, solutions proposed by Elallaoui et al. \cite{ELALLAOUI201842}, Gilson et al. \cite{GilsonFabian}, Avdeenko et al. \cite{AvdeenkoT19}, Muter et al. \cite{MuterLaurens}, etc., also utilized NLP techniques to handle ambiguities in user stories. However, they were lightweight (e.g., rule-based text identification); more advanced NLP methods can be utilized for further enhancement. 

Additionally, the above-mentioned frameworks have been proposed to identify ambiguities of user stories at a general level; they are not explicitly designed to identify privacy-related ambiguities. Considering privacy aspects of user stories, privacy-related ambiguities may play a crucial role, similar to the location example above. However, some of the ambiguity types may not be relevant when considering privacy aspects of user stories. For example, semantic ambiguities may not affect privacy considerations. Therefore, to develop the ambiguity detection step of our framework, we suggest adopting only the privacy-related ambiguity identification criteria from the above frameworks and tools and improving them. For example, the guidelines proposed by the QUS framework \cite{LucassenGarm15QUS} for developers to manually identify syntactic ambiguities can be automated (only if syntactic ambiguities affect privacy considerations). However, first we need to identify relevant ambiguities for our purpose. To do that, we recommend manually analyzing a set of user stories with the help of privacy experts to determine which types of ambiguities could impact the generation of privacy descriptions. 



Furthermore, from a technological standpoint, Berry et al. \cite{BerryDaniel12} suggested achieving 100\% recall (i.e., a perfect recall condition or zero false negatives) when identifying ambiguities, even by reducing the precision if necessary \cite{LucassenGarm16AQUSA}. That is because it ensures that the manual rechecking of ambiguities by going through the user story is not required. Considering all of these factors, we suggest utilizing more advanced ML models to replace the techniques used in the above methods. For example, to identify vague terms (i.e., they can be taken as either of four ambiguity types \cite{AMNA2022106824}) in user stories, we can feed user stories to a classification head on top of BERT's output to identify token/sentence-wise vagueness, since we have already processed user stories and converted them to embedding vectors in Step 3. This may require annotated data (i.e., a set of user stories and a set of vague terms) for training and validation purposes, which can be obtained using existing datasets \cite{RobeerMarcel16, CasilloFrancesco22, huang2020vague, neodataset2020} and some surveys with software development companies. 

Finally, if ambiguities are found, developers will be notified (e.g., the vague terms will be highlighted) through an interface, and they must correct them. 


\subsubsection{Generate privacy descriptions}
\label{priv_desc}

Once the ambiguities have been resolved, the next step (i.e., step 5) is to generate GDPR background information relevant to the user story, which is suggested to be achieved by training an LLM specifically for this purpose. In the context of GDPR, it is important to understand that data protection principles may evolve over time because of changes in legal standards, technological advancements, or emerging privacy concerns. Therefore, we propose developing this with an external, updatable knowledge base known as Knowledge Graph (KG) \cite{kg21}, which results in a KG-Augmented LLM \cite{PanShirui24}. 

First, we need to create a structured GDPR Knowledge Graph (KG) (e.g., Neo4j database \cite{neo4j2025}) that maps user actions (e.g., "delete account") and entities (e.g., medical records) found in user stories to relevant GDPR articles (e.g., Article 17) and the GDPR article requirements (e.g., "Right to Erasure"). To achieve this, we need to create triplets (i.e., the building blocks used to represent information in knowledge graphs) that map GDPR articles and their corresponding information to elements of user stories (e.g., user actions). This process can be supported by leveraging both privacy experts and NLP techniques. For example, as an NLP technique, we can use the Semantic Role Labeling (SRL) model \cite{srl_Marquez} to identify the 'who,' 'what,' and 'why' elements of a user story based on a given verb (which can be extracted using the spaCy\footnote{https://spacy.io/} library). These elements can then be mapped to relevant GDPR articles with the assistance of privacy experts. This process helps to make triplets such as ["user", "performs", "delete account"], ["delete account", "requires\_compliance", "GDPR Article 17"], and ["GDPR Article 17", is\_about, "Right to Erasure"]. 



Next, to fine-tune an LLM (e.g., LLaMA) for generating comprehensive descriptions, we need to create a database containing triplets retrieved from the KG (e.g., using a Cypher query \cite{neo4j_cypher_manual}) that is relevant to the elements (e.g., actions and entities) in user stories, as well as include pre-written descriptions related to those elements. For instance, the database may contain an entry such as: "\{User story: As a user, I want to upload my passport for identity verification. Extracted data: (Passport, is, Personal\_Data), (Personal\_Data, defined\_in, Article\_4\_1), (Purpose\_Limitation, described\_in, Article\_5\_1\_b). Description: This user story involves the processing of personal data, which is classified under Article\_4\_1...\}." After the database has been prepared, we need to use a decoder model (e.g., LLaMA) and train it to generate descriptions since we have the BERT embeddings that came from a previous step (i.e., Step 4). Then, we suggest applying LoRA (Low-Rank Adaptation) \cite{hu2021loralowrankadaptationlarge} to efficiently train the LLM for this particular case, as it reduces the training cost (e.g., number of trainable parameters). Hugging Face Transformers (i.e., to access LLMs) \cite{HuggingFace_Transformers} and the spaCy library (i.e., for NER) can be utilized to build this model. 

Once trained, at the inference stage, the output of Step 4 (i.e., the preprocessed user story) should be fed into the model to extract key elements from the user stories (using the same techniques discussed earlier). This should be followed by querying the KG to retrieve related GDPR information based on the extracted elements, and finally, using this information as input to the description generation LLM.

\subsubsection{Real-world scenario suggestion}
\label{real-world}
The next step of the framework is to suggest real-world (i.e., already occurred) GDPR noncompliance scenarios to developers by relating them with user story contexts to motivate developers to implement GDPR requirements. First, we need to create a database of historical real-world GDPR noncompliance cases. The database can be created by scraping online sources such as GDPR Enforcement Tracker \cite{enforcementtracker2025}, European Data Protection Board \cite{edpb2025}, news articles, and technical blogs. Second, we suggest creating a dataset by manually mapping each identified noncompliance case with the relevant GDPR article (i.e., violated article) with the help of privacy experts. 

At the inference stage, we can use the GDPR data (i.e., triplets) retrieved from the knowledge graph in the previous step (Step 5) to identify the relevant GDPR articles. Then, we can use the previously created dataset (i.e., the dataset that maps real-world incidents to GDPR articles) to suggest related real-world GDPR noncompliance cases based on the identified GDPR articles. 


\subsubsection{Privacy attitude evaluation}
\label{priv_attitude_eval}

To assess the privacy attitudes of software developers (i.e., to check whether the attitude has improved or not), we propose a questionnaire that should be answered on a Likert scale. To create questions, we suggest following the Theory of Planned Behavior (TPB) framework \cite{AJZEN1991179} that discusses three core components that drive human behavior: attitudes (i.e., how a person feels about the behavior), subjective norms (i.e., about social pressure or what they think others expect them to do), and perceived behavioral control (i.e., how easy or hard a person thinks it is to perform the behavior). Here, behavior refers to an action or response performed by a person. 

To develop the questionnaire, we propose a multi-step approach. First, qualitative methods such as individual developer interviews, group interviews (e.g., software teams in SMEs), and privacy expert discussions will be conducted to gather detailed information about developer behaviors corresponding to the three TPB components. For instance, we can ask a developer whether they believe that protecting user data is a valuable part of their development work, whether they feel their team expects them to follow GDPR requirements, and whether they feel confident in their ability to ensure GDPR compliance during development. 

Second, based on the qualitative findings, a preliminary pool of questions can be created that covers developers' attitudes, subjective norms, and perceived behavioral control. For example, "I believe that implementing privacy-preserving features adds value to the software I develop," "My team expects me to consider GDPR requirements when developing new features," and "I feel confident in my ability to identify and address potential GDPR compliance issues during development" are a set of questions corresponding to the three TPB components—attitudes, subjective norms, and perceived behavioral control, respectively—and should be answered using a Likert scale.

Third, the initial set of questions will be evaluated through developer feedback sessions where a small group of developers will be asked to complete the questionnaire and provide feedback on their clarity, relevance, and comprehensiveness. Fourth, questions can be refined further based on the given feedback. 

Finally, a pilot study will be conducted with a broader sample of developers from different regions (e.g., Europe, Asia, etc.) to assess the reliability statistically. Here, since the questionnaire is designed to assess developers' privacy attitudes, we will develop a tool based on the framework (similar to Figure \ref{fig-usecase} in Abstract \ref{sec-apndx-usecase}) before the pilot study. Then, during the study, developers will first be asked to complete the questionnaire, then use the tool, and finally complete the questionnaire again to evaluate how their attitudes have changed.

\section{A use case scenario}
\label{usecase}

To describe how the framework works, let's consider a hypothetical tool (e.g., a Jira plugin \cite{atlassian_sdk_setup, atlassian_jira_rest_api}) developed by incorporating the elements of the proposed framework (i.e., the discussed elements under \ref{prop_fw}) and explain how developers should interact with it to obtain the final output from the framework, as illustrated in Figure \ref{fig-usecase}. 

Consider the following scenario: A software developer who follows the Agile methodology specifies functional requirements as user stories for an application, and (s)he uses a Jira plugin that either can detect user stories specified in the Jira board or upload the entire set of user stories as a document (Figure \ref{fig-usecase}, top-left). Either way, the tool ends up with a list of functional requirements in the form of user stories, as can be seen under "Functional User Stories" in Figure \ref{fig-usecase} - top-center. Once uploaded, the tool will automatically correct the grammar and spelling issues as well as the format of the developer-written user stories. For example, as shown in Figure \ref{fig-usecase} - top-center, the identified grammar and format errors are corrected, and the developer is provided feedback to manually recheck and refine further adjustments. The "pen" icon can be used to refine the user story based on the given feedback. Once the spelling, grammar, and format errors are resolved, the developer can proceed to the next step, which is the ambiguity identification stage. In that stage, the tool will detect the ambiguities associated with each user story, notify the developer about them, and ask the developer to correct them and recheck the ambiguities if necessary. For example, as shown in Figure \ref{fig-usecase} - top-right, the term "medical records" is ambiguous, which is identified by the tool, and asks the developer to refine the story. Figure \ref{fig-usecase} - bottom-left shows the refined version of the user story once the identified ambiguities are corrected by the developer. Once the ambiguities are resolved, the tool will generate comprehensive GDPR-based privacy descriptions related to the user story to enhance the developer's GDPR awareness. It contains how to comply with the GDPR when implementing the functionality specified in the user story, why that GDPR compliance is needed, and which part of the GDPR compliance it was extracted from. An example of a generated privacy description is shown in Figure \ref{fig-usecase} - bottom-left (\textbf{Bold} for how to comply with GDPR, \textit{Italics} for why that compliance is needed, and \underline{Underline} for which part of the GDPR it was extracted from). Additionally, as shown in Figure \ref{fig-usecase}- bottom-left, if a privacy requirement is not required for a particular story, it will be indicated to the developer. Furthermore, to motivate developers to apply the gathered GDPR awareness through privacy descriptions (i.e., to comply with GDPR requirements) in practice, the framework provides real-world consequences of GDPR violations, as they will provide them a sense of what would happen if they could not comply with those GDPR requirements. As an example, when the developer clicks on the "see more" hyperlink next to the privacy description, a pop-up screen will appear that contains the real-world consequences as shown in Figure \ref{fig-usecase}, bottom-center. Finally, as shown in the bottom right of Figure \ref{fig-usecase}, developers are provided with a questionnaire form to evaluate their privacy attitude level by answering a set of Likert scale questions at any time. Developers can check their privacy attitude level at any time by clicking on the "Check" button on the top section of the interface, which also shows the level of privacy attitude.  

\section{Conclusion and future work}
\label{concl_fw}

In this paper, we propose a novel framework called
"GDPRShield" for software developers in SMEs to improve their GDPR awareness. The GDPRShield leverages functional requirements written as user stories to provide tailored GDPR-related privacy descriptions for each user story. These descriptions articulate the user story's relevance to GDPR articles and outline how to comply with relevant GDPR articles. By providing comprehensive GDPR-based privacy descriptions, the framework aims to enhance GDPR awareness among software developers. Since there is a correlation between software developers' privacy awareness and privacy attitudes, as we argued in the introduction, the GDPR-based privacy descriptions that the framework provides will enhance the developers' privacy attitudes while enhancing their GDPR awareness. In parallel, the framework provides real-world GDPR non-compliance scenarios, linking them to the context of the user stories. Since there are benefits for SMEs for complying with GDPR, as we discussed in the introduction, and seeing the real-world non-compliance scenarios (e.g., financial penalties), we argue that the software developers in SMEs are motivated to comply with GDPR. Overall, the "GDPRShield" is designed to improve the GDPR awareness of software developers in SMEs (i.e., those who do not have privacy expertise within their organization) through motivation, which eventually affects the software developers' privacy attitudes and the overall privacy culture of the organization. 

Future work will focus on developing a developer-centered tool (i.e., developed by incorporating the elements of the proposed framework) guided by Nielsen’s usability criteria for user interface design for "GDPRShield" \cite{nielsen_10usability_1994}. This tool will be rigorously evaluated using Nielsen’s usability criteria \cite{nielsen_5_usability} to ensure it is effective, efficient, and satisfying for developers. As a practical implementation, we plan to develop a Jira extension (plugin) by seamlessly integrating the framework functionalities into developers’ existing workflows, providing clear, actionable guidance without disrupting their productivity.

Also, we plan to conduct a further evaluation process of the framework through the tool among developers in SMEs in different regions (e.g., Europe, Asia, etc.) around the world, since the region can influence the level of GDPR awareness among software developers \cite{AlhazmiAbdulrahman21, PrybyloMaxwell24}. Also, different scenarios in the development process will be taken into account, such as situations where developers in SMEs face tight deadlines that may lead them to neglect privacy requirements, which are conditions that should also be evaluated.

The GDPRShield relies on machine learning models for different purposes (e.g., privacy description generation). Therefore, we need to examine how we can be compliant with GDPR when training and inferring with machine learning models \cite{FeretzakisGeorgios25}. In the future, we will also focus on proposing a comprehensive set of guidelines that should be followed when training machine learning models.

In Section \ref{real-world}, we suggested creating a database of real-world GDPR non-compliant scenarios by scraping online sources. However, the number of non-compliant cases may continue to grow and is not fixed. Therefore, in the future, we will develop a public repository that any developer can access through a portal and update with non-compliant scenarios they discover from online sources.

GDPR is a highly influential and widely adopted privacy regulation globally, which is why we chose to provide GDPR awareness through the framework. Even though some other regulations, such as the California Consumer Privacy Act (CCPA) \cite{ccpa18}, the Australian Privacy Act \cite{apa88}, and the New Zealand Privacy Act \cite{nzpa20} have been passed by the governments, they were inspired by the GDPR, incorporating the GDPR's core principles \cite{LiZeWerner22}. However, the GDPR applies to organizations that process the data of European Union (EU) residents. Therefore, SMEs that do not handle EU data may be less interested in adopting the framework unless they have a strong interest in understanding GDPR. Future work can focus on extending this framework to support other regulations (e.g., the Australian Privacy Act, New Zealand Privacy Act, etc.) to scale the framework in a way that is suitable for SMEs that deal with individuals outside Europe.

Finally, the ultimate goal of GDPRShield is to be a flexible and practical solution that helps software developers build more privacy-preserving systems, regardless of the regulation or the area in which the organization operates.

\bibliographystyle{IEEEtran}
\bibliography{IEEEabrv,IEEEfull}

\begin{appendices}

\section{Different types of ambiguities}
\label{sec-apndx-ambi}
The table (i.e., Table \ref{table:ambiguities}) describes the types of ambiguities that can be found in user stories \cite{LucassenGarm15QUS, AMNA2025112357, LucassenGarm16AQUSA, DALPIAZ20193, ELALLAOUI201842}.

\begin{table*}[h!]
\caption{Possible ambiguities present in user stories.}
\label{table:ambiguities}
\scriptsize
\centering
\begin{tabular}{|p{0.12\textwidth}|p{0.23\textwidth}|p{0.24\textwidth}|p{0.31\textwidth}|}
  \hline
  Type of ambiguity & Meaning & Example & Ambiguity terms \\ 
  \hline
  Lexical ambiguity & arises when words have different meanings and are specified with vague terminologies. & As a delivery driver, I want to see \textbf{user locations} so that I can complete deliveries efficiently. & The term "user locations" is underspecified, where it may be home, general area, or GPS location.\\
  \hline
  Syntactic ambiguity & arises when the grammatical structure of the user story leads to different interpretations. & As a patient, I need to be able to \textbf{enter my current medical records and personal information, and previous medical records can be edited} so that I can keep my profile up-to-date. & This makes different meanings, such as {enter my current medical records} and {personal information and previous medical records can be edited}, and {enter my current medical records and personal information} and {previous medical records can be edited}.\\ 
  \hline
  Semantic ambiguity & somewhat similar to lexical ambiguity. It arises when entire sentences or phrases have different interpretations because the words and their arrangements have multiple meanings. & As a patient, I want \textbf{enhanced health article suggestions}, so that I can read them. & Here, "enhanced" can be referred to as a number-wise or quality-wise improvement, but has not been clearly communicated. \\ 
  \hline
  Pragmatic ambiguity & arises when the context of the user story is interpreted differently. & As a doctor, I want to \textbf{view my patient’s medical records} so that I can comment on them. & Here, medical records can be viewed by exporting/downloading them as a file or using a separate interface within the dashboard. It has to be clearly specified.\\ 
  \hline
\end{tabular}
\end{table*}

\section{Use case scenario}
\label{sec-apndx-usecase}

\begin{figure*}[h]
    \centering
    \includegraphics[width=\textwidth]{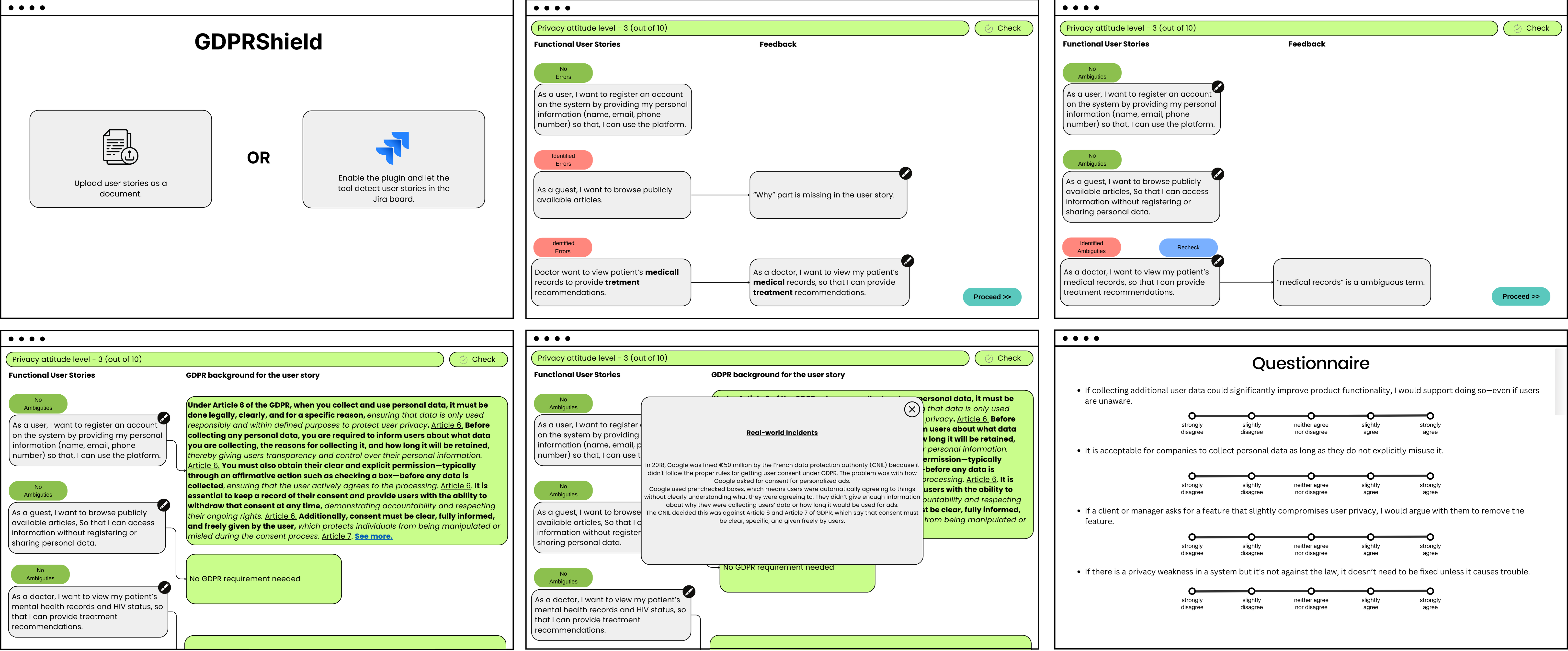}
    \caption{A set of wireframe views demonstrating the use case of the proposed framework built as a tool.}
    \label{fig-usecase}
\end{figure*}

Figure \ref{fig-usecase} shows simple wireframe views of a tool built incorporating the elements of the proposed framework.

\end{appendices}

\end{document}